\documentclass[12pt,preprint]{aastex}

\usepackage{bm}

\shorttitle{$N$-body + MHD Simulations of Merging Clusters of
       Galaxies}
\shortauthors{Takizawa}

\begin{document}


\title{$N$-body + Magnetohydrodynamical Simulations of Merging Clusters of
       Galaxies: Characteristic Magnetic Field Structures Generated by
       Bulk Flow Motion}


\author{Motokazu Takizawa}
\affil{Department of Physics, Yamagata University, 
       Yamagata 990-8560, Japan}
\email{takizawa@sci.kj.yamagata-u.ac.jp}



\begin{abstract}
 We present results from $N$-body + magnetohydrodynamical simulations of
 merging clusters of galaxies. We find that cluster mergers cause 
 various characteristic magnetic field structures because of the
 strong bulk flows in the intracluster medium. 
 The moving substructures result in cool regions
 surrounded by the magnetic field. These will be recognized as magnetized
 cold fronts in the observational point of view. A relatively ordered
 magnetic field structure is generated just behind the moving
 substructure. Eddy-like field configurations are also formed by
 Kelvin-Helmholtz instabilities. These features are similarly seen even in
 off-center mergers though the detailed structures change slightly.
 The above-mentioned characteristic magnetic field structures are partly
 recognized in Faraday rotation measure maps. The higher absolute values
 of the rotation measure are expected when observed along the collision
 axis, because of the elongated density distribution and relatively ordered
 field structure along the axis.
 The rotation measure maps on the cosmic microwave background
 radiation, which covers clusters entirely, could be useful probes
 of not only the magnetic field structures but also the internal
 dynamics of the intracluster medium.
\end{abstract}



\keywords{galaxies: clusters: general --- magnetic fields ---
          magnetohydrodynamics: MHD --- X-rays: galaxies: clusters }


\section{Introduction}
Clusters of galaxies have plenty of hot plasma as well as galaxies and dark
matter (DM), which is called intracluster medium (ICM). Some observational
evidence indicates that ICM is magnetized. For example, some
clusters have diffuse non-thermal synchrotron radio emission that is
called radio halos or relics, which shows that there are magnetic
fields as well as relativistic electrons in the intracluster space
\citep{Giov99, Kemp01}.  In addition, 
Faraday rotation measure observations of polarized radio
sources such as radio lobes and AGNs behind and/or in clusters indicate
that the magnetic field structures are quite random \citep{Clar01, Vogt03,
Govo06}.

Comparing the synchrotron radio flux with the hard
X-ray one (or its upper limit) due to the inverse Compton scatterings of
cosmic microwave background (CMB) photons, we are able to estimate the
volume averaged magnetic field strength (or its lower limit). 
Typically, the strength of $\sim 0.1 \mu$G is obtained 
with this method \citep{Fusc99, Fusc05} 
though those detections of non-thermal hard
X-ray components are still controversial \citep{Ross04, Fusc07}.
On the other hand, somewhat higher values 
($\sim$ a several $\mu$G) tend to be obtained with the Faraday rotation measure
method \citep{Clar01, Vogt03, Govo06}
though these results depend on the detailed magnetic field structures  
that are not fully understood \citep{Enss03, Murg04}.

Although the magnetic energy density is typically $\sim $ a few percents
of the thermal one in the intracluster space, it is believed that the magnetic
fields play a crucial role in various aspects of ICM. 
It is likely that non-thermal particles are accelerated via shocks 
\citep{Sara99, TakiNait00, ToKi00, Mini01, Ryu03, Taki03, Inou05} 
and/or turbulence \citep{Rola81, Schl87, Blas00, Ohno02, Fuji03, Brun04},
where the magnetic fields play a definitive role in most theoretical
models of the particle acceleration.
Heat conduction is probably depends on
the magnetic field configurations because charged particles cannot freely
move in the direction perpendicular to the field lines. 
As we wrote before, magnetic field seems to be a minor component in global ICM
dynamics. However, it is possibly important in relatively
small scales, where fluid instabilities such as Rayleigh-Taylor and 
Kelvin-Helmholtz ones might be suppressed by magnetic tension.

Cluster mergers have a significant impact on ICM magnetic field evolution.
Turbulent motion excited by mergers would amplify the field strength
via dynamo mechanism. Moving substructures are expected to sweep the
field lines and form the cold subclumps surrounded by the field lines
\citep{Vikh01}. 
\citet{Asai04} and \citet{Asai07} performed two and three dimensional
magnetohydrodynamical(MHD) simulations of moving cold
subclumps in hot ICM in rather idealized situations, respectively,
which confirmed that expectation. This field structure might be
responsible for the suppression of heat conduction and fluid
instabilities, which is essential for the maintenance of cold
fronts.

A lot of numerical simulations about merging clusters have
been done, most of which are $N$-body + hydrodynamical simulations
\citep{Roet96, Taki99, Taki00, Rick01, Ritc02, Asca06, Taki06, 
McCa07, Spri07}. 
Although these simulations give us a great deal of understanding
of the structures, evolution, and observational implications for merging
clusters of galaxies, they make only a limited contribution to
investigate the magnetic field structures. 
MHD simulations are essential in this regard.
However, $N$-body + MHD simulations are rather rare though
\citet{Roet99} did pioneering work. 
It is true that Lagrangian particle methods based on smoothed particle
hydrodynamics \citep[see][]{Mona92} are extensively used in cosmological
MHD simulations \citep{Dola99, Dola02}. 
Considering that Eulerian mesh codes are essentially better at following
evolution of magnetic fields, simulations based on such codes are highly
desirable.
In this paper, we present the results from $N$-body + MHD simulations of
merging clusters of galaxies
and investigate characteristic magnetic field
structures during mergers and their implications.

The rest of this paper is organized as follows. In \S \ref{s:simu} we
describe the adopted numerical methods and initial conditions for our
simulations. In \S \ref{s:resu} we present the results. In \S
\ref{s:summ} we summarize the results and discuss their implications.

\section{The Simulations}
\label{s:simu}

\subsection{Numerical Methods}
Numerical methods used here are basically similar to those in
\citet{Taki06} except that ideal MHD equations are calculated for ICM
instead of ordinary hydrodynamical ones.
In the present study, we consider clusters of galaxies consisting of two
components: collisionless particles corresponding to the galaxies and
DM, and magnetized gas corresponding to the ICM. When calculating gravity, both
components are considered, although the former dominates over the
latter. Radiative cooling and heat conduction are not included.  
In calculating the MHD equations for the
ICM, we use a simple linearized Riemann solver originally proposed by
\citet{Brio88} and introduced into astrophysical MHD simulations by 
\citet{Ryu95}. This Riemann solver is a similar and simpler version of 
Roe's method \citep{Roe81}.
Using the MUSCLE approach and a minmod TVD limiter \citep[see][]{Hirs90},
we obtain second-order
accuracy without any numerical oscillations around discontinuities. To
avoid negative pressure, we solve the equations for the total energy and
entropy conservation simultaneously. This method is often used in
astrophysical hydrodynamic simulations where high Mach number flow can
occur \citep{Ryu93, Wada01}.

Gravitational forces are calculated by the
Particle-Mesh (PM) method with the standard Fast Fourier Transform
technique for the isolated boundary conditions \citep[see][]{Hock88}. 
The size of the simulation box is $(9.44 \,\mathrm{Mpc})^3$.
For typical runs, the number of the grid points is $(256)^3$,
and total number of the $N$-body particles is $(128)^3$, 
which is approximately $2.1 \times 10^6$.

\subsection{An Equilibrium Cluster Model}
We consider mergers of two virialized subclusters with an NFW density
profile \citep{Nava97} in the $\Lambda$CDM universe 
($\Omega_0=0.25$, $ \lambda_0=0.75$) for DM,
\begin{eqnarray}
  \rho_{\rm DM}(r) = \frac{\delta_{\rm c} \rho_{\rm crit}}
                   {(r/r_{\rm s})(1 + r/r_{\rm s})^2},
\end{eqnarray}
where $\delta_{\rm c}$, $\rho_{\rm crit}$, and $r_{\rm s}$ are 
the characteristic (dimensionless) density, critical density, and scale radius,
respectively.
Given the cosmological parameters and halo's virial mass, 
we calculate these parameters following a method in Appendix
of \citet{Nava97}.

The initial density profiles of the ICM are assumed to be
those of a beta-model,
\begin{eqnarray}
   \rho_{\rm gas}(r) = \rho_{\rm gas,0}  \biggl\{ 
               1 + \biggl( \frac{r}{r_{\rm c}} \biggr)   
                        \biggr\}^{-3 \beta /2},
   \label{eq:rhogas}
\end{eqnarray}
where $\rho_{\rm gas,0}$ and $r_{\rm c}$ are the central gas density and
core radius, respectively.
We assume that $\beta=0.6$, which is consistent with typical values
obtained form the recent X-ray observations \citep{Vikh06, Cros08}.
On the other hand, it is not easy to obtain the observational 
relationship between $r_{\rm c}$ and $r_{\rm s}$ considering difficulty 
of accurate determination of mass profile. Following the discussion in 
\citet{Rick01}, therefore, we chose $r_{\rm c}=r_{\rm s}/2$ as a fiducial 
value. The gas mass fraction is set to be 0.1 inside of the virial radius of 
each subcluster. It should be noted that the results will not be sensitive 
to the choice of this fraction unless the ICM dominates the DM in gravity.

The velocity distribution of
the DM particles is assumed to be an isotropic Maxwellian. The radial
profiles of the DM velocity dispersion are calculated from the Jeans
equation with spherical symmetry, so that the DM particles would be in
virial equilibrium in the cluster potential of the DM and ICM,
\begin{eqnarray}
   \frac{d}{dr}(\rho_{\rm DM} \sigma^2) = 
                  - \frac{G M(r)}{r^2} \rho_{\rm DM},
   \label{eq:jeans}
\end{eqnarray}
where $\sigma^2$ and $M(r)$ are the one dimensional velocity dispersion at
$r$ and total mass inside $r$, respectively. $G$ is the gravitational
constant. We need boundary conditions to obtain $\sigma^2(r)$ from
differential equation (\ref{eq:jeans}), which is assume to be 
\begin{eqnarray}
  \sigma^2(r_{\rm vir}) = \frac{G M(r_{\rm vir})}{3 r_{\rm vir}},
\end{eqnarray}
where $r_{\rm vir}$ is the virial radius of the halo.

The radial profiles of the ICM pressure are determined in a similar
way so that the ICM would be in hydrostatic equilibrium within 
the cluster potential,
\begin{eqnarray}
  \frac{dP}{dr} = - \frac{G M(r)}{r^2}\rho_{\rm gas}.
   \label{eq:hyst}
\end{eqnarray}
The boundary conditions for equation (\ref{eq:hyst}) is
\begin{eqnarray}
   P(r_{\rm vir}) = \frac{1}{\beta_{\rm spec}} 
                    \frac{G M(r_{\rm vir})}{3 r_{\rm vir}}
                    \rho_{\rm gas}(r_{\rm vir}),
\end{eqnarray}
where $\beta_{\rm spec}$ is the specific energy ratio of the DM and gas
at the virial radius.
In case of the classical isothermal $\beta$-model \citep[see][]{Sara86}, 
$\beta_{\rm spec}$ should be equal to the fitting parameter $\beta$ in equation
(\ref{eq:rhogas}) to obtain isothermal temperature profile.
Although our model is different than the isothermal $\beta$ model 
in underlying DM distribution, we assume $\beta_{\rm spec}$=$\beta$ 
for simplicity.
As a result, the temperature profile is not isothermal.

\subsection{Initial Conditions}

For typical runs, the DM masses within the virial radius of the larger
and smaller subclusters are
$5.0 \times 10^{14} \,\mbox{\ensuremath{{M}_{\odot}}}$ and 
$1.25 \times 10^{14} \,\mbox{\ensuremath{{M}_{\odot}}}$, 
respectively. Thus, the mass ratio
is $4:1$. It is useful to introduce an angular momentum parameter 
$\lambda \equiv J |E|^{1/2} / (G M^{5/2})$ 
in order to characterize off-center collisions, 
where $J$, $E$, and $M$
are the angular momentum of the two subclusters around the center of
masses, the binding energy between the two, and the total mass, respectively. 
Please note that 
$\lambda$ is the ratio between the actual angular velocity and the
angular velocity needed to provide rotational support
\citep[see][]{Binn87}.

Given each subcluster's mass and an angular momentum parameter of the system,
we estimate ``typical'' initial conditions for cluster mergers 
in a similar way of \citet{Taki99}. How to estimate these
conditions is described in detail in Appendix, which is a natural generalization
of the method for head-on collisions \citep[see][section 2]{Taki99} 
to cases including off-center collisions.
The coordinate system is
taken in such a way that the center of mass is at rest at the origin.
Two subclusters are initialized in the $xy$-plane, separated by a
distance $\sqrt{(R_1+R_2)^2-b^2}$ in the $x$-direction, 
and $b$ in the $y$-direction, 
where $R_{1,2}$ and $b$ are the virial radii of the each subcluster
and the impact parameter, respectively.
The initial relative velocity is directed along the $x$-axis. 
The centers of the larger and smaller subclusters were initially
located at the sides of $x<0$ and $x>0$, respectively. 

At present, we have only limited information about magnetic field
configurations in the intracluster space. Roughly speaking, however, 
they have random structures whose power spectrum can be approximated by
a power-law in a smaller scale, and the mean field strength
is an increasing function of the gas density.
Taking these features into account, we make the initial magnetic field as
follows. First, we generate random Gaussian vector potential in the
wave-numbers space with a zero mean and single power-law spectrum, 
$A(k) \propto k^{-\mu}$. We assume $\mu=5/3$.
Then, the generated potential is transformed into the real space 
using a three-dimensional Fast Fourier Transform and 
multiplied by a factor of $\rho_{\rm gas}^{2/3}$,
which is expected assuming a uniform spherical collapse and flux
freezing. A random and divergence-free magnetic field is calculated
via $\bm{B} = \bm{\nabla} \times \bm{A}$.
The normalization of the magnetic field is determined so that the total
magnetic energy is one percent of the thermal one. Strictly speaking,
each subcluster in the initial conditions is not in hydrostatic equilibrium
with magnetic field.  However, it does not a matter in our simulations
because the magnetic energy is much less than the thermal one 
as we wrote above.

The parameters for each model are summarized in table \ref{tab:modpar},
where $M_{i}$, $r_{{\rm s},i}$, $c_{i}$, $N_{i}$ are the total mass, 
scale radius, concentration parameter, and number of $N$-body particles 
for the $i$-th subcluster, respectively. $N_{\rm grd}$ is the total number of 
grid points. RunST, which is a simple head-on collision case 
with the mass ratio of 4:1, is regard as a standard run. RunOC is a case of 
an off-center merger with $\lambda=0.05$. RunMM is a relatively minor 
merger case with the mass ratio of 8:1. RunLR is a lower resolution run 
to investigate how the results depend on numerical resolution.

\section{Results}
\label{s:resu}
\begin{figure}
\includegraphics[angle=0,scale=0.85]{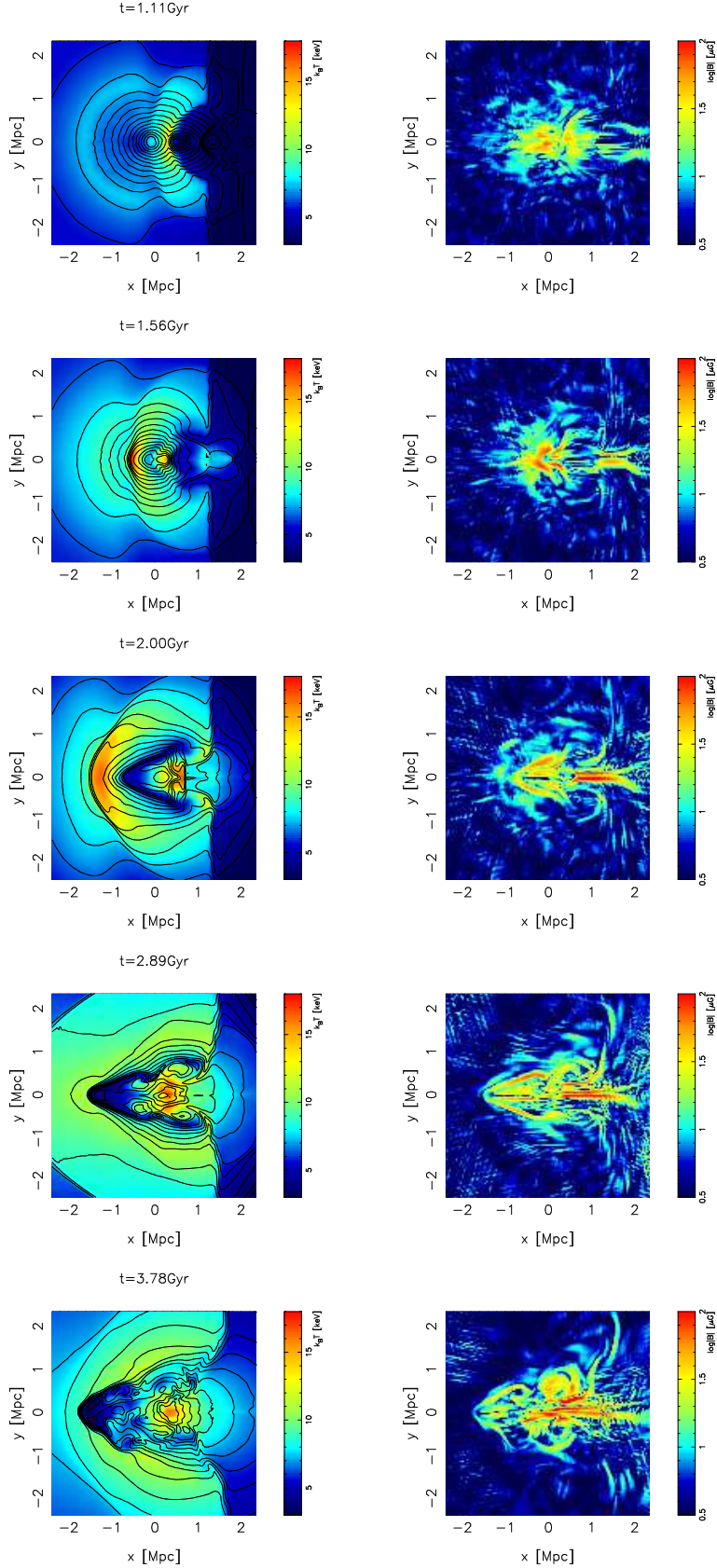}
\caption{Left: Snapshots of the density (contours) and temperature
 (colors) distributions on the  $z=0$ surface at $t=$1.11, 1.56, 2.00,
 2.89, and 3.78 Gyr of RunST. Right: Same, but for the
 magnetic field strength distribution.}
\label{fig:den_tem_mag}
\end{figure}

The left panels of Figure \ref{fig:den_tem_mag} show snapshots of the
density (contours) and temperature (colors) on the $xy$-plain, which is
perpendicular to the collision axis,  at $t=1.11$, $1.56$, $2.00$,
$2.89$, and $3.78$ Gyr of RunST.
Right ones show the same but for the magnetic field strength.
At $t=1.11$ Gyr, two subclusters are approaching each other, and
temperature between two density peaks is higher than elsewhere. However,
magnetic field strength in this region is lower than that in
the density peaks. This is simply because initial mean magnetic field
strength increase with the density. Although the magnetic
field there is slightly amplified by the adiabatic compression, this 
does not compensate the trend of the initial magnetic field.
At $t=1.56$ and $2.0$ Gyr, slightly amplified magnetic field 
is seen just behind the shock. However, two kinds of structure appear
more notably. One is a relatively strong magnetic field region
associated with a contact discontinuity between the ICM originated from
the smaller and larger subclusters, which will be recognized as a cold
front with strong magnetic field in the observational point of view. As
a result, a cool region surrounded by the magnetic field appears. 
The other is an ordered magnetic field in $x$-direction 
just behind the subcluster. This is because
flow behind subcluster is converging on the collision axis. 
As a result, the magnetic field are collected and amplified because of
compression.
At $t=2.89$ Gyr, the bow shock has gone outside of the panel. the field
structure associated with the contact discontinuity and ordered field
along the collision axis are still present. In addition, eddy-like
field structures start to appear above and below the ordered field along 
the collision axis in $xy$-plain. 
These structures grow as Kelvin-Helmholtz instabilities
develop and become clearer at $t=3.78$ Gyr

\begin{figure}
\includegraphics[angle=0,scale=0.85]{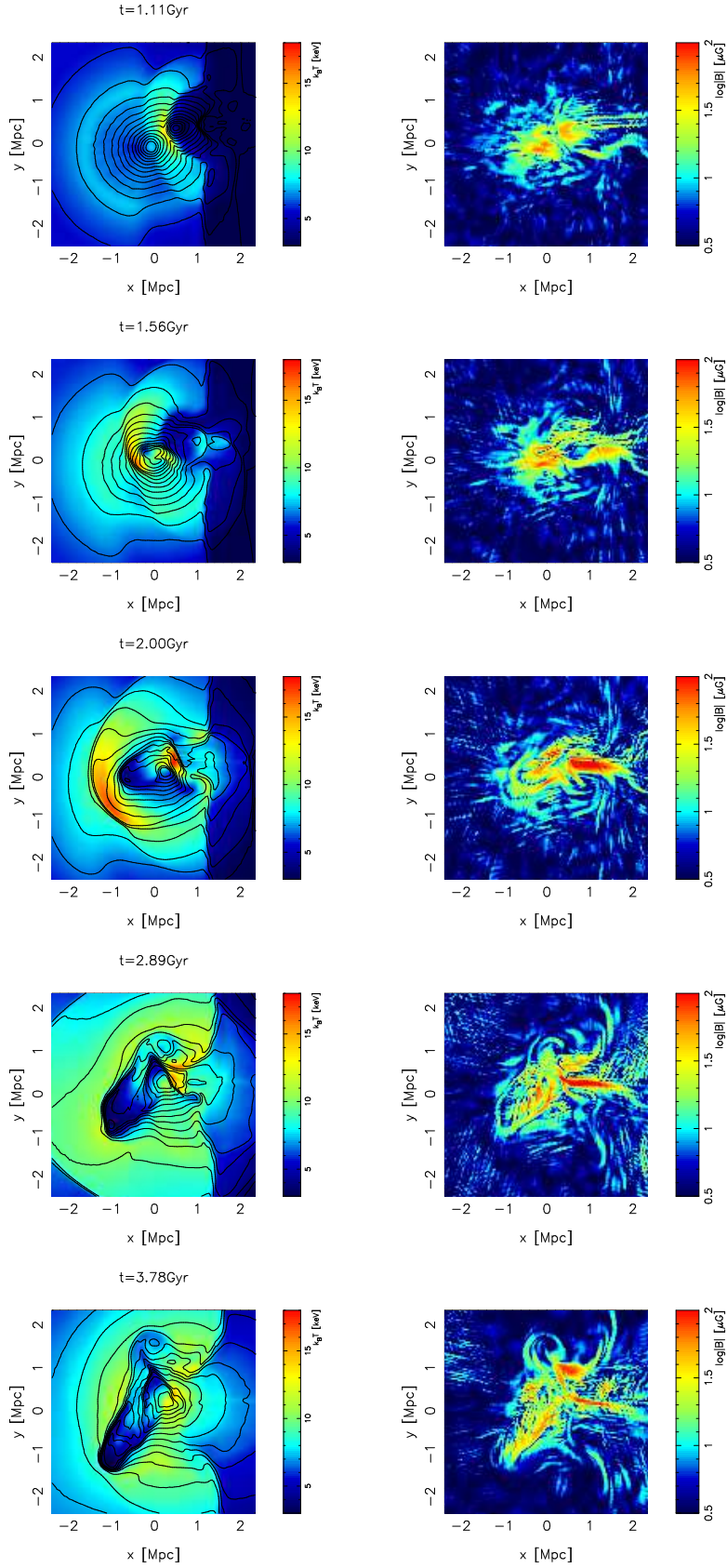}
\caption{Same as Fig. \ref{fig:den_tem_mag}, but for RunOC.}
\label{fig:den_tem_mag_2}
\end{figure}

The initial conditions of RunST have an axial symmetric
structure. Thus, the above-mentioned features could change significantly
in off-center mergers.
Figure \ref{fig:den_tem_mag_2} shows the same as figure \ref{fig:den_tem_mag},
but for RunOC.
Again, the prominent magnetic structures are associated with a contact
discontinuity rather than a bow shock.
Because of the asymmetry, the magnetic field at the contact discontinuity
is stronger in the side closer to the lager cluster core (or lower side
in each panel) at $t=2.00$ and $2.89$ Gyr. A cool region surrounded by
the magnetic field certainly appears, but the field structure in the side
farther to the larger cluster core (or upper side in each panel)
becomes less clear at $t=3.78$ Gyr.
As in the case of RunST, an ordered and relatively strong
magnetic field structure appears behind the moving substructure. 
However, this is not just behind the clump but shifted towards the lager
cluster side.
Eddy-like magnetic fields generated by the Kelvin-Helmholtz instabilities
are clearer in the side further to the larger cluster.

\begin{figure}
\includegraphics[angle=0,scale=0.85]{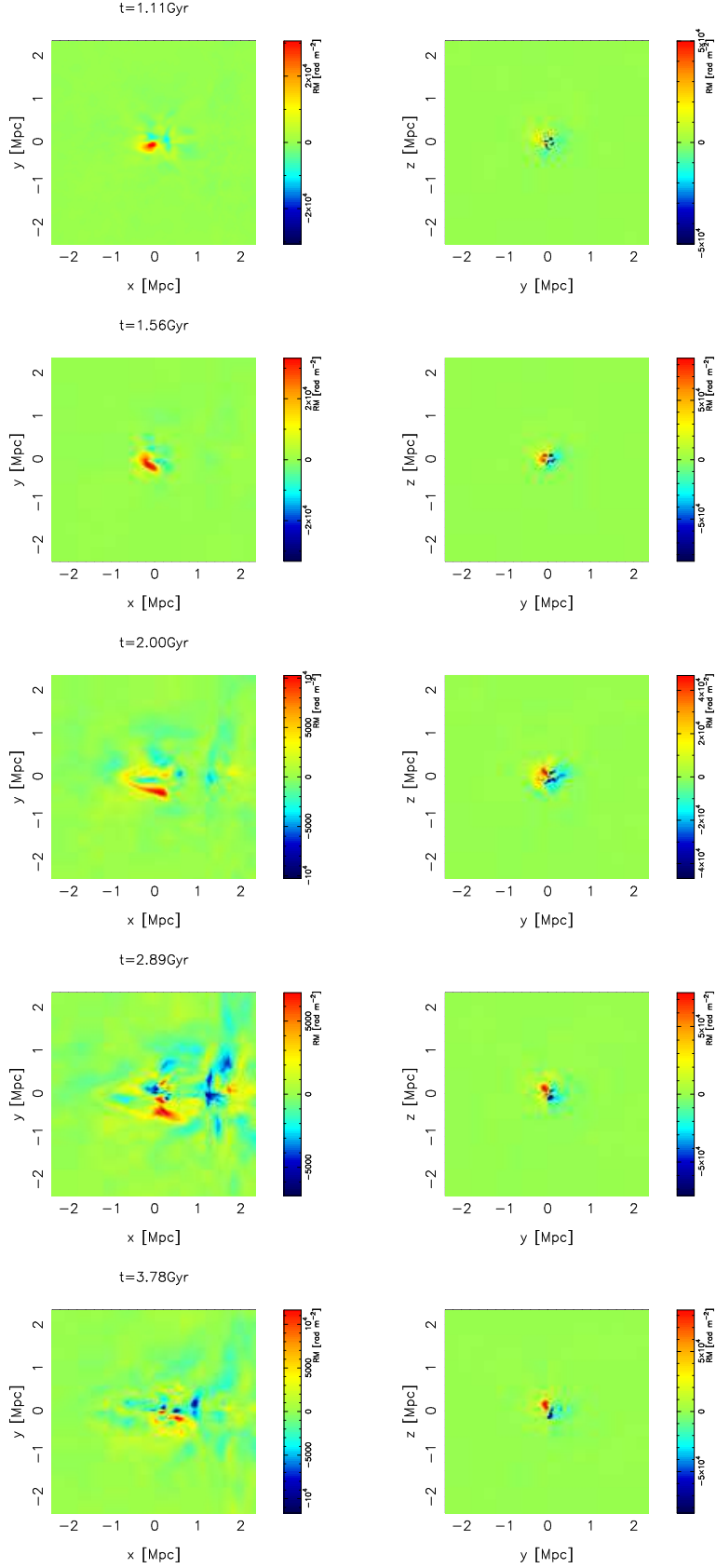}
\caption{Snapshots of the Faraday rotation measure map of RunST
   at the same epochs as in Fig \ref{fig:den_tem_mag}. Left and right
   panels show the rotation measure when the line-of-sight is
   perpendicular and parallel to the collision axis, respectively.
   }
\label{fig:frm}
\end{figure}

Cluster mergers cause various characteristic magnetic field configurations
as we see before.
Unfortunately, it is very difficult to observe the magnetic field
structure itself directly at present. The observation of Faraday
rotation measure is an indirect, but quite useful way to obtain
information of the magnetic field structure.
The rotation measure (RM) is given by \citep[see][]{RyLi79},
\begin{eqnarray}
  {\rm RM} [{\rm rad}/{\rm m}^{2}] = 812 \int n_{\rm e}[{\rm cm}^{-3}]
                              B_{||}[\mu{\rm G}] dl,
  \label{eq:frm}
\end{eqnarray}
where $n_{\rm e}$ and $B_{||}$ are the electron number density and 
line-of-sight component of the magnetic field, respectively.
Figure \ref{fig:frm} presents snapshots of the RM map for RunST.
Left and right panels show the maps seen from the $z$ and $x$ axis,
respectively. Please remember that the collision axis is along the $x$-axis.
When the line-of-sight is perpendicular to the collision axis, RM
maps give us relatively rich information.
Cool regions with high magnetic field are seen as those with high
absolute values of RM. On the other hand, the structure is relatively
featureless when we observe the system along the
collision-axis. The absolute values of RM tend to be higher in the
right panels, which is not surprising because the density distribution 
is elongated along the collision axis, and because there is the ordered
magnetic field in the same direction behind the subclump.

\begin{figure}
\includegraphics[angle=270,scale=0.65]{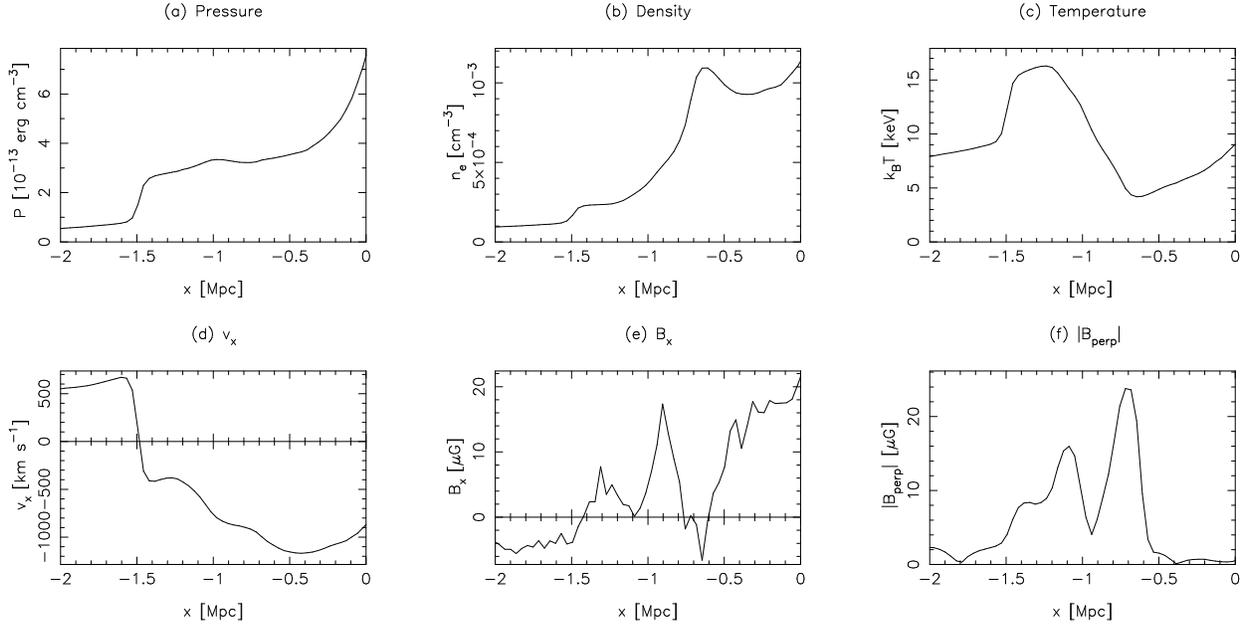}
\caption{Profiles along the $x$-axis at $t=2.00$ Gyr of RunST,
 for (a)pressure, (b)electron number
 density, (c)temperature, (d)$x$-component of velocity, (e)$x$-component
 of magnetic field, and (d)absolute value of magnetic field
 perpendicular to the $x$-axis, respectively.}
\label{fig:axisplot}
\end{figure}

Figure $\ref{fig:axisplot}$ shows profiles along the collision axis at
$t=2.00$ Gyr of RunST for (a) pressure, (b) electron number
density, (c) temperature, (d) $x$-component of velocity, (e)
$x$-component of the magnetic field, and (d) absolute value of the
magnetic field perpendicular to the axis. There is a bow shock at 
$x \simeq -1.5$, where a clear discontinuity are seen in the profiles of
pressure, density, temperature, and $x$-component of velocity.
A somewhat blunt contact discontinuity is at $x \simeq -1$, where density and
temperature have a jump, but pressure and $x$-velocity are smooth.
The magnetic field perpendicular to the axis also increase across the
bow shock. The tangential magnetic field is amplified between the shock
and the contact discontinuity. As a result, the cool region is wrapped by
the magnetized layers. This characteristic structure probably have an
influence on the transport process such as heat conduction, which will
be discussed later.

\begin{figure}
\includegraphics[angle=270,scale=0.65]{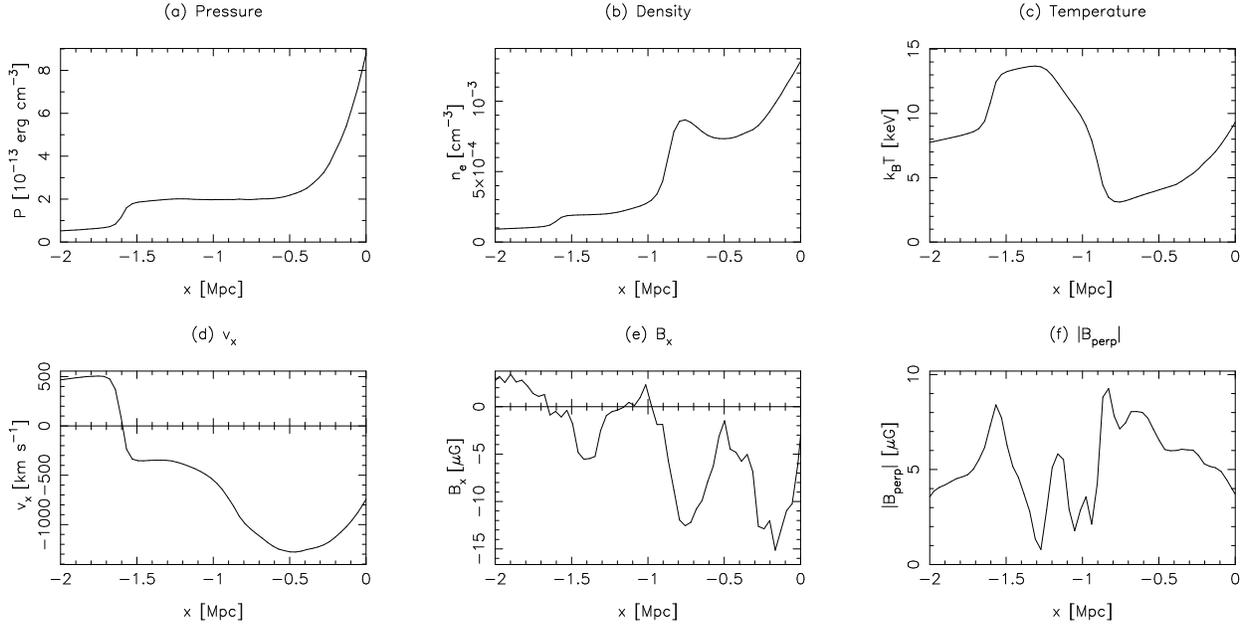}
\caption{Same as Fig. \ref{fig:axisplot}, but for RunMM.}
\label{fig:axisplot_2}
\end{figure}

Because minor mergers are more frequent in actual situations, it is interesting
how the results change in minor mergers. Figure \ref{fig:axisplot_2} shows
the same as figure \ref{fig:axisplot} but for RunMM, where mass ratio is 1:8.
Mach number of the bow shock becomes lower, which is 
clearly seen in the profile of pressure, density, temperature, and $x$-component
of velocity.
As a result, amplification of the magnetic field component perpendicular to the 
collision axis around contact discontinuity
is less prominent, where typical values are roughly 
halves of those in RunST. On the other hand, typical strength of the 
$x$-component of the magnetic field does
not change very much.

\begin{figure}
\includegraphics[angle=270,scale=0.65]{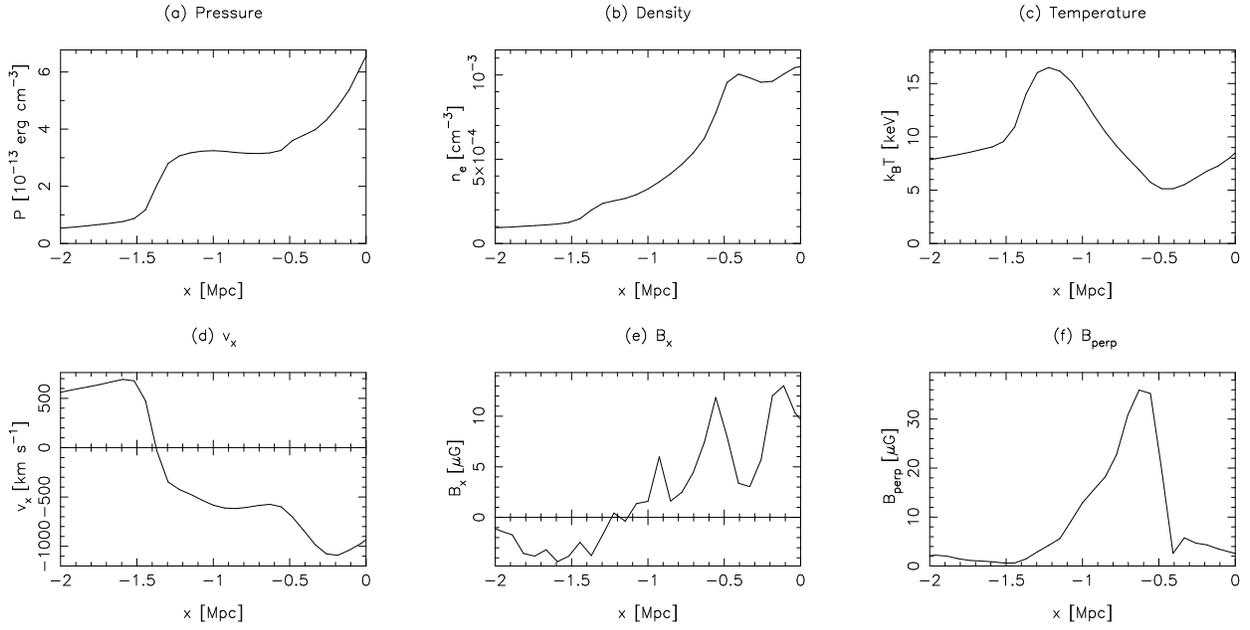}
\caption{Same as Fig. \ref{fig:axisplot}, but for RunLR.}
\label{fig:axisplot_3}
\end{figure}

It is possible that our results, especially on the small scale structures 
around the cold front and bow shock, depend on numerical resolution. 
To assess this, we perform a lower resolution run as RunLR. 
Both the numbers of grid points and particles in RunLR 
are eighths of those in RunST. The initial conditions of RunLR is 
essentially the same as those of RunST 
though the largest wave-number components 
in the magnetic field are not included.
Figure \ref{fig:axisplot_3} shows the same as figure \ref{fig:axisplot} but for
RunLR. Although the overall features in pressure, density, temperature, and 
$x$-component of velocity
are similar to those of RunST qualitatively, it is obvious that the width of 
the contact discontinuity is broader. Width of the bow shock is also slightly 
broader. As for magnetic field, although the small scale structures 
disappear, amplification of the field component perpendicular 
to the collision axis near the contact discontinuity
is clearly seen again.

\section{Summary and Discussion}
\label{s:summ}
We perform $N$-body and MHD simulations of merging clusters of
galaxies. We find that cluster mergers cause various characteristic magnetic
field structures because of strong bulk flow motion. 
The magnetic field component perpendicular to the
collision axis is amplified especially between a bow shock and contact
discontinuity. As a result, a cool region wrapped by the field lines
appears. A relatively ordered field structure along the collision axis
appears just behind the moving substructure. Eddy-like field structures
are also generated by Kelvin-Helmholtz instabilities. Although the
detailed structures change slightly, similar features are seen in
off-center mergers.
RM maps have some information about the
magnetic field structure. The above-mentioned characteristic structures
are partly recognized in the RM maps when we observe the merging systems
in the direction perpendicular to the collision axis. On the other hand,
RM maps observed nearly along the collision axis are less informative
in this respect.
Typical absolute values of the RM become higher when the system is
observed along the
collision axis because of both the density distribution elongated toward
the axis and ordered magnetic field along the same direction.
In minor mergers, amplification of the magnetic field component 
perpendicular to the collision axis becomes less prominent.
The results about magnetic field along the collision axis does not change 
very much. Although numerical resolution effects have the significant impact on
the small scale structures such as widths of a bow shock
and contact discontinuity and detailed small scale fluctuations of the magnetic
field, our results about overall global structures are reasonably reliable.

Some observational and theoretical studies suggest that heat conduction
in the ICM is suppressed from the Spitzer value.
For example, a sharp temperature jump across a cold front in A2142 
requires that the heat conductivity is reduced by a factor of between
250 and 2500 at least in the direction across the front \citep{Etto00}.
The spatial temperature variations in the central region of A754
also suggest that the conductivity is at least an order of magnitude lower
than the Spitzer value \citep{Mark03}. 
It is well-known that fine-tuning of heat conductivity is necessary
in order to reproduce observational natures of cool
cores assuming that radiative cooling in the cores balances with the
conductive heating from the outer part of the clusters \citep{Zama03}.
Although the detailed process involved there is still unclear, magnetic
fields likely play an important role. Heat conduction in the direction
across the magnetic field lines is likely suppressed because electrons cannot
travel freely in that direction. Using two dimensional MHD simulations 
with anisotropic heat conduction, \citet{Asai04} shows that a moving
subclump naturally form the magnetic field structure along the contact
discontinuity and that the temperature jump there is maintained if the
heat conduction perpendicular to the field line is sufficiently suppressed. 
Similar results are obtained in three dimensional cases
\citep{Asai07}. Basically, our simulations confirmed their results about
magnetic field configurations in more realistic situations, 
though the heat conduction is not included.

We show that Faraday rotation measure maps are useful to obtain
information about characteristic magnetic field structures formed by 
cluster mergers. This could be another probe of internal dynamics of
clusters. However, there is a serious problem in this method at
present. We need polarized radio sources such as radio lobes and/or AGNs
in and/or behind the clusters of galaxies to do that. In other words, 
we are able to obtain RM maps in the limited regions where we have 
suitable polarized sources by chance. Naturally, these do not always
correspond to the regions that we are interested in. However, this difficulty 
could be overcame if we have suitable polarized sources that covers a cluster
entirely. One possible solution is to use the CMB as the 
polarized source \citep{Ohno03},
though this is still challenging in the present status
of CMB observations. 
There are a lot of observations that are ongoing and planned for
measuring the CMB polarization. We hope that future observations of the CMB 
polarizations will enable us to make RM maps that cover clusters
entirely, which would give us an important clues to understand the
internal dynamics as well as the magnetic field structures in clusters
of galaxies.

In actual clusters, the magnetic field have random components in small
scales. It is very difficult to treat such scales by numerical
simulations that concern the global structures.
Clearly, observed RM maps are influenced by the small scale magnetic field
fluctuations, which are not considered in RM maps calculated from our
results. As a result, coherent length of
the magnetic fields tends to be overestimated effectively, which means
absolute values of RM from the simulations are also overestimated. However,
it is probable that global spatial patterns of RM maps are nearly 
independent of such small scale fluctuations. In addition, it is
also probably robust that higher RM are expected when the merging
systems are observed along the collision axis.



\acknowledgments

The author would like to thank N. Asai for valuable discussion.
The author is also grateful to an anonymous referee 
for his/her useful comments and suggestions
which improved the manuscript.
Numerical computations were carried out
on VPP5000 at the Center for Computational Astrophysics, CfCA, of the
National Astronomical Observatory of Japan and PRIMEPOWER 850 at
Networking and Computing Service Center of Yamagata University.
M. T. was supported in part by a Grant-in-Aid from the
Ministry of Education, Science, Sports, and Culture of Japan (16740105,
19740096).





\appendix
\section{An Estimation Method of the Initial Conditions of Mergers}
The estimation method introduced here is a natural extension of that for head-on
collisions in \citet{Taki99} to cases including off-center collisions.
We consider a merger of two subclusters with masses $M_1$ and $M_2$
($M_1 \ge M_2$).
Let us consider a region that contains both subclusters. It expands
following the cosmological expansion at first. However, the expansion is
decelerated and then the region will collapse if the mean density is
higher than the critical one. 
We assume that they are separated by the distance of $2r_{\rm ta}$
at the maximum expansion epoch. In case of head-on collisions,
it is obvious that they are at rest for their center of mass at that
time. In case of off-center mergers, on the other hand, they have
the relative tangential velocity $v_0$ at that time.
Taking into account the conservation of energy and angular momentum between 
at the maximum expansion epoch and just before the merger, we obtain
following equations,
\begin{eqnarray}
  - \frac{G M_1 M_2}{2 r_{\rm ta}} + \frac{1}{2} M v_0^2  &=&
  - \frac{G M_1 M_2}{R_1 + R_2} + \frac{1}{2} M v^2, 
    \label{eq:en_con}           \\
  2 M v_0 r_{\rm ta} &=& M v b 
    \label{eq:angm_con},
\end{eqnarray}
where $R_1$, $R_2$, $v$, and $b$ are the virial radius for each
subcluster, the initial collision velocity, and the initial impact
parameter, respectively. $M \equiv M_1 M_2 /(M_1 + M_2)$ is the reduced
mass of the system. 
It is natural that the virial radius has correlation with the
mass. According to a spherical collapse model, the virial mass is
proportional to the square of the virial radius \citep[see][]{Peeb80}.
Thus, we assume that the following relation,
\begin{eqnarray}
   R_2 = \biggl( \frac{M_2} {M_1} \biggr)^{1/3} R_1 .
   \label{eq:r2_r1}
\end{eqnarray}
The spherical collapse model also tell us that the turn around radius
is twice of the virial radius. Thus, we also assume,
\begin{eqnarray}
    r_{\rm ta} = 2 \biggl( \frac{M_1 + M_2}{M_1} \biggr)^{1/3} R_1 .
    \label{eq:rta_r1}
\end{eqnarray}
Using equations (\ref{eq:en_con}), (\ref{eq:angm_con}),
(\ref{eq:r2_r1}), and (\ref{eq:rta_r1}), we eliminate $v_0$ and obtain
$v$ as follows,

\begin{eqnarray}
  v^2 = \frac{2GM_1}{R_1} ( 1 + \alpha )
         \biggl\{
       \frac{1}{1+\alpha^{1/3}} - \frac{1}{4(1+\alpha)^{1/3}}
         \biggr\}
         \biggl\{
       1 - \frac{1}{16(1+\alpha)^{2/3}} \biggl(\frac{b}{R_1}\biggr)^2
         \biggr\}^{-1}, \label{eq:v2} \\
\end{eqnarray}
where
\begin{eqnarray}
   \alpha \equiv \frac{M_2}{M_1} \label{eq:alpha} .
\end{eqnarray}
It is useful to introduce an angular momentum parameter 
$\lambda \equiv J |E|^{1/2} / (G M^{5/2})$ in order 
to characterize off-center mergers, where $J$, $E$, and $M$
are the angular momentum of the two subclusters around the center of
masses, the binding energy between the two, and the total mass,
respectively. Please note that 
$\lambda$ is the ratio between the actual angular velocity
and the angular velocity needed to provide rotational support
\citep[see][]{Binn87}.

From equations (\ref{eq:en_con}) and (\ref{eq:angm_con}),
we obtain
\begin{eqnarray}
  \lambda = \frac{v b}{(G M_1 R_1)^{1/2}} 
            \frac{\alpha^{3/2}}{(1+\alpha)^{7/2}}
            \biggl\{
              \frac{1}{1+\alpha^{1/3}} - 
              \frac{R_1 v^2}{2GM_1(1+\alpha)}
            \biggr\}^{1/2}. \label{eq:lambda}
\end{eqnarray}
Given $M_1$, $R_1$, $\alpha$, and $\lambda$, therefore, we are able to
calculate $v$ and $b$ from equations (\ref{eq:v2}) and
(\ref{eq:lambda}).




\clearpage



\clearpage
 \begin{table}
  \caption{Model Parameters}
  \label{tab:modpar}
  \begin{center}
   \begin{tabular}{ccccc}
    \hline \hline 
  Parameters   & RunST    &  RunOC  &  RunMM   &  RunLR \\
    \hline
 $M_1/M_2 (10^{14} \mbox{\ensuremath{{M}_{\odot}}})$ & 5.0/1.25      & 5.0/1.25  & 5.0/0.625  &  5.0/1.25 \\
 $r_{{\rm s},1}/r_{{\rm s},2}$ (kpc)                  & 256/139       & 256/139   & 256/104    &  256/139 \\
 $c_1/c_2$                                         &  6.12/7.08     & 6.12/7.08    &  6.12/7.56     & 6.12/7.08 \\
 $\lambda$                                         &  0.0          & 0.05       &  0.0     &   0.0\\
 $N_1/N_2$                                     &  1677721/419431  & 1677721/419431  & 1864135/233017 & 209715/52429  \\
 $N_{\rm grd}$                       &  $(256)^3$       & $(256)^3$    & $(256)^3$  & $(128)^3$  \\
    \hline
   \end{tabular}
  \end{center}

 \end{table}%

\end{document}